# Electrostatic model for antenna signal generation from dust impacts

Mitchell M. Shen[1,2], Zoltan Sternovsky[1,2], Alessandro Garzelli[1,3], David M. Malaspina[1,4]

1. *Laboratory for Atmospheric and Space Physics, University of Colorado, Boulder, CO 80303*
2. *Smead Aerospace Engineering Sciences Department, University of Colorado, Boulder, CO 80303*
3. *Dipartimento di Scienze e Tecnologie Aerospaziali, Politécnico di Milano, Via La Masa, MI 20156, Italy*
4. *Department of Astrophysics and Planetary Sciences, University of Colorado, Boulder, CO 80309*

**Key Points:**

(1) Analysis and interpretation of dust impact signals are improved by a new electrostatic model based on recollected and induced charge effects.

(2) A laboratory model is constructed to verify the validity of presented model and measure the basic properties of dust impact plasma plumes.

(3) The electrostatic model provides the framework for interpreting antenna waveforms and determining the parameters of impacting dust particles.

*Manuscript submitted to JGR*

***Corresponding author:*** uniecoaass@gmail.com

# Abstract

Dust impacts on spacecraft are commonly detected by antenna instruments as transient voltage perturbations. The signal waveform is generated by the interaction between the impact-generated plasma cloud and the elements of the antenna – spacecraft system. A general electrostatic model is presented that includes the two key elements of the interaction, namely the charge recollected from the impact plasma by the spacecraft and the fraction electrons and cations that escape to infinity. The clouds of escaping electrons and cations generate induced signals, and their vastly different escape speeds are responsible for the characteristic shape of the waveforms. The induced signals are modeled numerically for the geometry of the system and the location of the impact. The model employs a Maxwell capacitance matrix to keep track of the mutual interaction between the elements of the system. A new reduced-size model spacecraft is constructed for laboratory measurements using the dust accelerator facility. The model spacecraft is equipped with four antennas: two operating in a monopole mode, and one pair configured as a dipole. Submicron-sized iron dust particles accelerated to > 20 km/s are used for test measurements, where the waveforms of each antenna are recorded. The electrostatic model provides a remarkably good fit to the data using only a handful of physical fitting parameters, such as the escape speeds of electrons and cations. The presented general model provides the framework for analyzing antenna waveforms and is applicable for a range of space missions investigating the distribution of dust particles in relevant environments.

# 1. Introduction

Antenna instruments on space missions are primarily designed to measure electric fields and plasma waves [*Gurnett*, 1998]. These instruments have also been found sensitive to the detection of cosmic dust particles as they impact the surface of the spacecraft or the antennas [*Gurnett et al.,* 1983, 1997; *Kurth et al.,* 2006; *Meyer-Vernet et al.*, 2009, 2017; *Malaspina et al.*, 2014, 2020; *Ye et al.,* 2014, 2016a, 2016b, 2018, 2019; *Kellog et al.,* 2016; *Szalay et al.,* 2020; *Vaverka et al.,* 2018, 2019; *Zaslavsky et al.,* 2012, 2015].

Dust impacts are detected as transient voltage signals generated by the expanding impact plasma cloud. Impact ionization is a physical phenomenon, where free charges in the form of electrons, cations, and anions are generated upon the high-speed impact ($\gtrsim 1$ km/s) of dust particles on solid surfaces [e.g., *Auer*, 2001]. The generated impact charge ($Q_{IMP}$) scales approximately linearly with the dust particle's mass ($m$), increases steeply with speed, and depends on the target material, as determined from laboratory measurements [e.g., *Auer*, 2001; *Collette et al.,* 2014 and references therein]. The transient voltage signals (waveforms) measured by the antennas can be used to obtain valuable information on the impacting dust particles as mentioned above. Dust impacts can also be detected as wideband 'noise' in the antenna signals' power spectrum [*Aubier et al.,* 1983; *Meyer-Vernet et al.,* 2009]. This article is limited to the interpretation of the waveforms from individual impacts.

Antenna instruments may operate in monopole or dipole mode, where the voltage difference between an antenna and the spacecraft (SC) or between two antennas is measured, respectively. The antenna and SC acquire equilibrium potentials in space due to charging currents from photoelectron emission and the collection of electrons and ions from the ambient plasma. The equilibrium potential can be positive or negative (depending on the relative magnitude of the charging currents), and it will affect the expansion of the impact ionization plasma cloud. The dust impacts are registered as transient voltage perturbations imposed on top of the equilibrium potentials then will relax back to the equilibrium values over timescales which are characteristic of the environment.

Several models have attempted to describe the physical mechanisms leading to the generation of voltage signals measured by antennas. *Gurnett et al.* [1983, 1987] assumed that a significant fraction ($\alpha$) of the impact charge electrons is collected by the antennas with positive equilibrium potentials. The voltage measured by the antenna is then $V = \alpha Q_{IMP}/C_A$, where $C_A$ is the capacitance of the antenna. In addition, *Gurnett et al.* [1987] considered the collection of the impact charge by the SC in the dipole mode. In this case, a small fraction of the SC voltage is measurable by the antennas, given as $V = \gamma Q_{IMP}/C_{SC}$, where $C_{SC}$ is the SC capacitance, and $\gamma$ is the common-mode rejection coefficient of the electronics. In theoretical work, *Oberc* [1996] identified three mechanisms that can lead to the generation of voltage signals: charging of the antenna, charging of the SC, and sensing of the charge separation electric field. The first two mechanisms are similar to the models described above. In the third one, the antennas are sensing the electric field of the ion cloud during the expansion of the impact plasma. The model assumes that the impact plasma is moving away from the impact location and simultaneously is expanding over time. At some point during the expansion, the electrons are decoupled from the plasma cloud, leaving behind the cloud of slower ions with a positive space charge potential on the order of the electron temperature expressed in the units of eV. The antennas then detect the separated electric field of the cloud. *Oberc* [1996] has pointed out that the measured signals generated by antenna charging would also strongly depend on the impact geometry, i.e., the impact location relative to the geometry of the antennas and the SC. On the other hand, signals generated by spacecraft charging would be independent of the impact geometry. *Oberc* [1996] noted that for monopole antennas, the dominating mechanism is due to SC charging. For dipole antennas, the measured signals are from the differential charging of the two antennas and sensing the separated electric field. The relative importance between the latter two mechanisms depends on the physical characteristics of the antennas, such as their lengths. *Zaslavsky* [2015] proposed the floating potential perturbation model for the data collected by the STEREO/WAVES instrument [*Bale et al.,* 2008; *Bougeret et al.,* 2008]. This model assumes that both the SC and the antenna recollect some fraction of the impact plasma. The signal measured by the antenna is then the difference of the voltage perturbations on the antenna and the SC due to charge collection. The final characteristic shapes of the measured waveforms are set by the different discharge time-constants of the two elements through the ambient plasma. *Meyer-Vernet et al.* [2017] proposed an analytical model for calculating the risetime of antenna signals. This work pointed out several additional key aspects of the signal generation mechanisms. For example, the electrons in the impact plasma acquire an isotropic velocity distribution due to their high thermal speed; thus, half of the electrons move toward the spacecraft rather than away from it after charge separation. In addition, charge $Q$ at a small distance from the surface of the SC will induce a potential with magnitude $\sim Q/C_{SC}$, i.e., it has a similar effect as the same charge collected on the SC. *Kellogg et al.* [2018] noted that there is capacitive coupling between the SC and the antenna and the base resistor installed in between these elements need to be considered in determining the discharging time constant.

The investigation of the antenna signal generation processes in laboratory conditions was made possible by the dust accelerator facility at the University of Colorado [*Shu et al.,* 2012]. Using a simple setup, *Collette et al.* [2015] identified three different signal generation mechanisms, namely SC charging, antenna charging, and induced charging. The polarity of the SC and antenna charging signals can be reversed by changing the polarities of the applied bias voltages on the elements. Numerical analysis has later shown that charge collection by the antennas is effective only for dust impacts occurring in the close vicinity of the antenna base [*O'Shea et al.,* 2017]. For typical impacts analyzed for the STEREO SC and its antenna instruments, the collection efficiencies of the antenna themselves are only on the order of $0.1 - 1$ %. *Nouzák et al.* [2018, 2020] performed laboratory studies using a 20:1 scaled-down model of the Cassini SC and the Radio and Plasma Wave Science (RPWS) instrument, in both monopole and dipole modes. The measured waveforms in the laboratory are in good qualitative agreement with those measured in space and demonstrated how the waveform features vary with bias voltage on the SC. The measurements also confirmed that antennas in a dipole mode are insensitive to dust impacts on the SC body, and that of the magnetic field affects the recollection of electrons from the impact plasma considering the gyro motions.

Auxiliary measurements were also performed for characterizing the impact charge yields of various materials [*Collette et al.*, 2014 and references therein] and the effective temperatures of the electrons and ions in the impact plasma. The electron temperatures were found to be on the order of $1 - 4$ eV, and ion temperatures between $4 - 15$ eV, increasing with impact speed for iron (Fe) dust particles impacting on a tungsten (W) target [*Collette et al.* 2016; *Nouzák et al.,* 2020]. Measurements using olivine dust particles with an organic coating indicated about 7 eV ion temperatures and slight variation with impact speed. The negative charge carriers' temperature varied over $1 - 10$ eV with non-monotonic velocity dependence over the $3 - 18$ km/s range [*Kočiščák et al.,* 2020].

The lab measurements collectively lead to a refined *qualitative* physical model for the generation of antenna signals that can be summarized as follows: The impact plasma, consisting of electrons and ions, can be divided into fractions recollected by the spacecraft or escape. The ratio of collected/escaping fractions of electrons and ions is determined by the SC potential and the effective temperatures of the respective species. The two main signal generation mechanisms are SC charging from the net recollected charge, and induced charging from the escaping fraction. The characteristic waveforms are generated in four successive and somewhat overlapping steps. First, the fast escape of the electrons leaves behind a net positive charge on and near the SC, generating a steep, negative-going signal measured as $V_{ANT}(t) - V_{SC}(t)$. This feature is known as the 'preshoot', which is commonly observed by antenna instruments that operate with sufficiently wide bandwidths and adequate sampling rates [see, for example, *O'Shea et al.,* 2017]. Second, the escape of electrons is followed by that of ions, driving the signal more positive. Third, once the charge escape is completed, the SC is left with the net collected charge that is responsible for the main peak in the waveform. And fourth, the voltages on the SC and the antennas relax back to their equilibrium values as the system is discharging through the ambient environment. This discharge process operates through the duration of the event and may significantly reduce the amplitude of the main peak [*Shen et al.*, 2021]. The time constant of the discharge process is set by the magnitudes of charging currents from the environment [*Zaslavsky*, 2015]. The description of the overall antenna signal generation processes is also provided in a review article by *Mann et al.* [2019].

*Shen et al.* [2021] have recently presented the *quantitative* analytical form of the model described above. It is applied on a simplified system consisting of one antenna and a spherical SC. The latter assumption allows expressing the induced charge on the SC in a simple analytical form, $Q_{SC,ind}(t) = Q_{esc}\frac{R_{SC}}{R_{SC}+r(t)}$. Here $R_{SC}$ is the SC radius, $r(t) = vt$ is the radial distance of the charge escaping with velocity $v$, and the escaping charge $Q_{esc}$ is approximated as a point charge moving radially outward [*Jackson*, 1999]. The quantitative model provides good fits to the waveforms collected using a model SC. Several fundamental parameters of the impact plasma cloud are determined by fitting the model to the data, including the impact charge ($Q_{IMP}$), ion expansion speed, etc.

This article expands upon the model presented by *Shen et al.* [2021] and generalizes it for an arbitrary geometry of the SC-antenna system. The model employs a capacitance matrix to calculate the voltages developed on the elements from the collected and induced charges. A new laboratory model SC has also been constructed to investigate the effects of the impact location on the waveforms. The model SC is spherical for simplicity and employs four antennas: two operated as monopoles and one dipole pair. The preliminary analysis of the dataset demonstrates that the electrostatic model can accurately describe waveforms measured in the laboratory using the dust accelerator.

The article is organized as follows: Section 2 described a generalized electrostatic model for the generation of antenna signals. The new experimental setup and the recipe for obtaining the capacitance matrix are described in Sec. 3. Section 4 describes the numerical fitting routine, which includes the effects of the sensing electronics, and provides the preliminary analysis of data for two impact geometries. The summary and conclusions are provided in Sec. 5.

# 2. Electrostatic Model

## 2.1 Collected and Escaping Charges

The model described below follows and expands upon that presented by *Shen et al.* [2021]. The generated impact charge has the form of a power law,

$$Q_{IMP} = Q_i = |Q_e| = \gamma \, m v^{\beta}, \tag{1}$$

where $m$ is the mass of the dust particle, and $v$ is the impact speed. Parameters $\gamma$ and $\beta$ are characteristics of the target material, and their values can be determined from laboratory measurements [e.g., *Auer*, 2001; *Collette et al.*, 2014]. In the first approximation, the impact plasma consists of electrons ($Q_e$) and cations ($Q_i$) of equal quantity. Here we make the same assumptions as of the previous work by *Shen et al.* [2021]: (a) The voltage signals due to the dust impacts are small compared to the equilibrium floating potential; thus, the impact signal can be treated as a perturbation. (b) The electrons and cations in the impact plasma decouple from one another in the expanding plasma cloud over a distance that is short compared to the characteristic size of the spacecraft ($R_{SC}$). Consequently, the electrons and cations can be treated independently, each expanding with their characteristic speed. These two assumptions limit the validity of the presented model in terms of the impact charge to $Q_{IMP} \lesssim 1 \times 10^{-9}$ C for typical SC dimensions [*Shen et al.*, 2021].

The electron and ion components of the impact plasma are divided into escaping and re-collected fractions. For a positively charged SC ($V_{SC} > 0$), these fractions are described as:

$$Q_{e,esc} = -\kappa \, Q_{IMP} \, e^{\frac{-V_{SC}}{T_e}}$$

$$Q_{e,col} = -Q_{IMP} - Q_{e,esc}$$

$$Q_{i,esc} = Q_{IMP}$$

$$Q_{i,col} = Q_{IMP} - Q_{i,esc} = 0 \tag{2}$$

where $T_e$ is the effective electron temperature in unit of eV. Equation (2) describes a physical picture, where the ions of the impact plasma are expanding away from impact location in form of a plume, and the faster electrons acquire an isotropic distribution in the early stages of the expansion. Coefficient $\kappa$ describes the unobstructed fraction of solid angle in which the isotropic electrons can escape. The value of $\kappa$ ranges from $0 - 1$, and for a semi-infinite impact surface $\kappa \approx$ 0.5, as half of the isotropic electrons will be moving away from the surface, while the other half towards the surface, and will be recollected. The escape of the electrons is also governed by energetics that is described by a Boltzmann factor. For negative SC potential ($V_{SC} < 0$), the governing equations become:

$$Q_{e,esc} = -\kappa \, Q_{IMP}$$

$$Q_{e,col} = -Q_{IMP} - Q_{e,esc}$$

$$Q_{i,esc} = Q_{IMP} \, e^{\frac{V_{SC}}{T_i}}$$

$$Q_{i,col} = Q_{IMP} - Q_{i,esc} \, , \tag{3}$$

where $T_i$ is the effective ion temperature in unit of eV.

## 2.2   Simplified Case with One Antenna

First, the simplified, yet illustrative example of a SC with one antenna is provided, considering only collected charges on one element. Figure 1 depicts this case, where $C_{SC}$ and $C_{ANT}$ are the physical capacitances of the respective elements. The mutual capacitance, $C_x$, consists of several components. It includes the 'base' capacitance ($C_B$), the capacitances of the cables and preamp input capacitance ($C_{WP}$), and the capacitance of the antenna to the SC body that is determined by the geometry of the system ($C_G$) [see, for example, *Bale et al.,* 2008]. Since these three capacitances are effectively connected in parallel, $C_X = C_B + C_{WP} + C_G$.

The next step is considering that charge $\delta Q_{SC,col}$ is deposited onto the SC. The notation '$\delta$' designates perturbation, i.e., the deposited charge is in addition to that responsible for the development of the equilibrium potential. The voltage that develops on the SC is then given by $\delta V_{SC} = \delta Q_{SC,col}/C_{SC,eff}$, where $C_{SC,eff} = C_{SC} + \frac{C_x C_{ANT}}{C_x + C_{ANT}}$ is the effective capacitance of the SC that includes the contributions from $C_{ANT}$ and $C_x$. Even though no charge is deposited onto the antenna, the voltage of these elements is affected by the voltage on the SC. This voltage is given as $\delta V_{ANT} = \delta V_{SC}[C_x/(C_x + C_{ANT})]$, following the rule of capacitors connected in series. The voltage measured between the antenna and the SC is then:

$$\delta V_{meas} = \delta V_{ANT} - \delta V_{SC} = -\frac{\delta Q_{SC,col}}{C_{SC,eff}} \frac{C_{ANT}}{C_x + C_{ANT}}. \tag{4}$$

The ratio at the end of Eq. (4) is usually considered as the antenna gain, $G_A = C_{ANT}/(C_x + C_{ANT})$ [see, for example, *Bale et al.,* 2008].

Similarly, for a charge deposited on the antenna, the measured voltage is:

$$\delta V_{meas} = \delta V_{ANT} - \delta V_{SC} = \frac{\delta Q_{ANT,col}}{C_{ANT,eff}} \frac{C_{SC}}{C_x + C_{SC}} \tag{5}$$

where $C_{ANT,eff} = C_{ANT} + \frac{C_x C_{SC}}{C_x + C_{SC}}$ is the effective capacitance of the antenna.

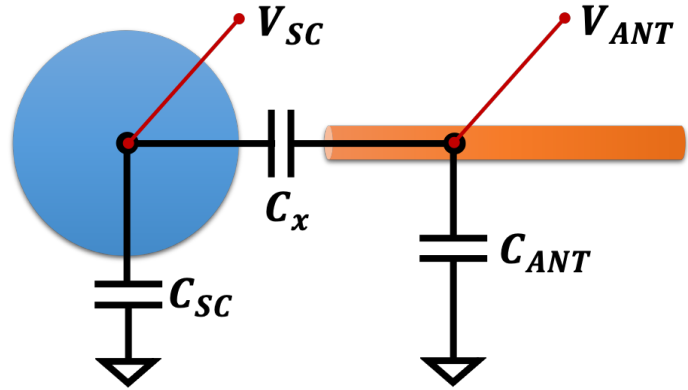

**Fig. 1:** Schematics of the simple system consisting of a spacecraft (SC) and one antenna (ANT). See text for details.

For the considered simplified cases, Equations (4) and (5) can be used to calculate the deposited charge from the amplitudes of the voltage signals. There are two noteworthy comments to make: One is that in the case of charge deposited onto the antenna, the antenna gain $G_A$ is no longer applicable. The second is that contrary to common practice in previous studies, the effective values of the SC and antenna capacitances need to be used to convert charge to voltage and vice versa. When the SC-antenna system is immersed in plasma, the capacitances of the elements will increase. This effect is, however, negligible for conditions where the Debye length of the plasma is longer than the characteristic size of the SC, e.g., in interplanetary space near 1 AU.

## 2.3   The Matrix Form and Induced Charging

The calculations of the effective capacitances and measured voltages increase complexity quickly with increasing the number of antennas. Fortunately, Maxwell's capacitance matrix can be employed to keep track of how the SC and the antenna elements interact [*Maxwell*, 1873; *Di Lorenzo*, 2011; *de Queiroz*, 2012]. For $n$ antennas, the matrix form can be written as

$$\begin{bmatrix} \delta V_{SC} \\ \vdots \\ \delta V_{ANT,n} \end{bmatrix} = \begin{bmatrix} a_{00} & \cdots & a_{0n} \\ \vdots & \ddots & \vdots \\ a_{n0} & \cdots & a_{nn} \end{bmatrix} \begin{bmatrix} \delta Q_{SC,col} \\ \vdots \\ \delta Q_{ANT,col,n} \end{bmatrix} \qquad (6)$$

which allows calculating the voltages on objects from known collected charges. In this form, $[\boldsymbol{a}]$ is the elastance matrix with $(n+1)^2$ elements. The inverse form to Eq. (6) is $[\delta Q] = [\boldsymbol{b}][\delta V]$, where $[\boldsymbol{b}] = [\boldsymbol{a}]^{-1}$ is the capacitance matrix. For the remainder of this section, it will be assumed that the elastance and capacitance matrices are known for the system. Section 3.2 below will provide the details of how the matrices can be calculated. In addition, Appendix A provides the details of the elastance and capacitance matrices for the simplest case of an SC with one antenna.

Induced charging refers to the fact that a free charge in the vicinity of a conductive object will induce a potential on this object. An analytical solution for the established potential exists for the simple case of a spherical object, where the magnitude of the induced potential scales as $1/r$ with radial distance from the surface (see Sec. 1 and *Jackson* [1999]). *Shen et al.* [2021] presented the analytical solution for this simple case with the SC in the spherical approximation.

The general solution for induced charging for the SC and the antenna elements can be calculated numerically, provided that the geometry of the system and the charge distribution are known. In order to describe the fraction of the charge induced on the *i*-th element from a test point charge ($Q_{test}$) located at position $\vec{r}$, the *geometric function* is introduced. For the system of an SC and *n* antennas, the geometric functions are defined as:

$$g_{SC}(\vec{r}) = \frac{Q_{SC,ind}(\vec{r})}{Q_{test}},$$

$$g_{ANT,n}(\vec{r}) = \frac{Q_{ANT,n,ind}(\vec{r})}{Q_{test}}, \tag{7}$$

where $Q_{SC,ind}(\vec{r})$, for example, is the charge induced on the SC, while the unit point charge is located at $\vec{r}$. The geometric functions provide unitless values and sum up to $g_{SC}(\vec{r}) + \sum_n g_{ANT,n}(\vec{r}) \leq 1$ for all locations. The value of the geometric function approaches unity for a point charge in the close proximity to one of the elements and diminishes with increasing distance.

The geometric functions can be calculated numerically using standard available electrostatic solver tools. The geometry of the SC – antenna system is imported into the software, and each element of interest is treated as a conductive object that is electrically isolated from the rest of the system. The software also calculates the elastance and/or capacitance matrices. The values of the geometric functions can be calculated simply from the voltages established on each element using the following relation:

$$\begin{bmatrix} g_{SC}(\vec{r}) \\ \vdots \\ g_{ANT,n}(\vec{r}) \end{bmatrix} = \frac{1}{Q_{test}} \begin{bmatrix} b_{00} & \dots & b_{0n} \\ \vdots & \ddots & \vdots \\ b_{n0} & \dots & b_{nn} \end{bmatrix} \begin{bmatrix} V_{SC} \\ \vdots \\ V_{ANT,n} \end{bmatrix}. \tag{8}$$

The calculation of the geometric function can be performed for all locations of interest in the vicinity of the system. The capacitance matrix in the equation above can account only for the geometric coupling between the elements, i.e., $C_x = C_G$. The description of how to account for the additional contributions from $C_B$ and $C_{WP}$ is described in Sec. 3.2. Given the size and complexity of an actual SC – antenna system and the required resolution, the numerical calculations of the geometric function can be numerically demanding. However, these calculations have to be performed only once and then use standard numerical techniques, e.g., lookup tables combined with linear interpolation for determining the values for arbitrary positions.

Figure 2 shows the numerical calculations for a simplified case of a spherical model SC with one antenna. The dimensions of the system resemble those of the model used in the laboratory measurements described in Sec. 3. The SC radius is $R_{SC} = 7.62$ cm, and the antenna is 27 cm long and 1.6 mm in diameter. In the model setup, there is a 5 mm gap between the surface of the SC and the antenna, which limits their mutual coupling. The point charge is moving along a radial line that is offset 10° from the antenna, as illustrated in the insert of Fig. 2. The magnitude of the test point charge is $Q_{test} = 100$ pC. The left panel shows the potentials on each element, and their difference, $V_{meas} = V_{ANT} - V_{SC}$. When the point charge is near the surface of the SC, the potentials are $V_{SC} \cong Q_P/C_{SC,eff} = 10.8\ V$, where $C_{SC,eff} = 9.60\ pF$ is the effective SC capacitance, as calculated. This is close to the expected value of $C_{SC,eff} = C_{SC} + \frac{C_x C_{ANT}}{C_x + C_{ANT}} = 9.46$ pF, where $C_{SC} = 4\pi\varepsilon_0 R_{SC} = 8.47$ pF is the capacitance of the SC, and the values of $C_x = 1.43$ pF and $C_{ANT} = 1.84$ pF are also from the numerical calculations. With the point charge near the surface of the SC, the potential of the antenna is $V_{ANT} \approx 4.6$ V, which is close to the value calculated as $V_{ANT} = V_{SC}[C_x/(C_x + C_{ANT})] = 4.7$ V. The minor discrepancies in the calculated and theoretical values are likely due to numerical errors (e.g., grid or numerical setup in the software).

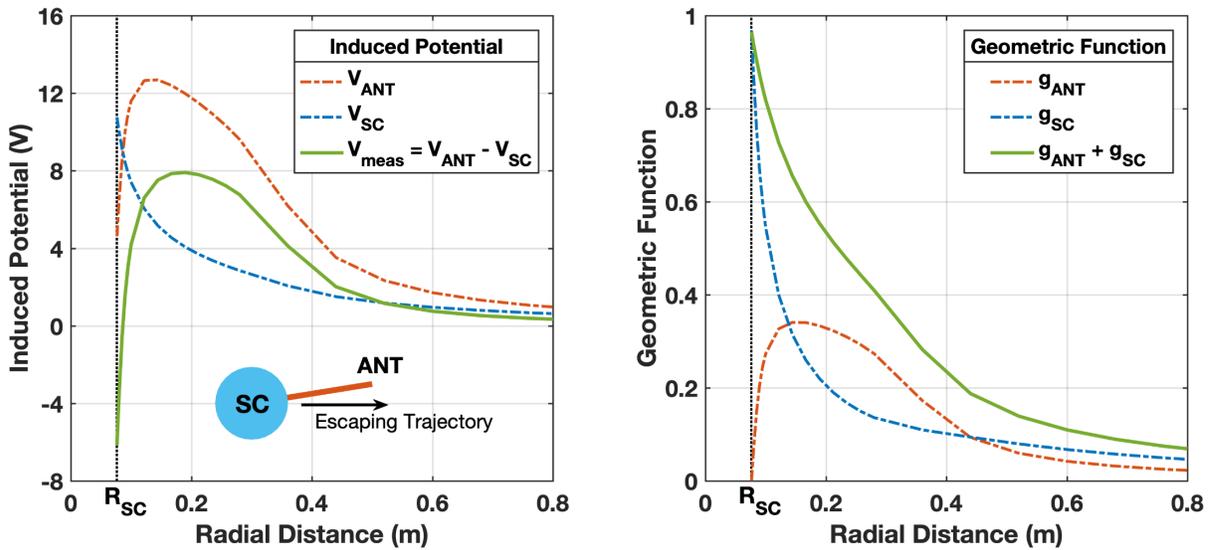

**Fig. 2:** The induced potential (left panel) and the geometric function (right panel) for a simple case of a system consisting of a spherical SC and one antenna. The numerical calculations are for a test point charge ($Q_{test} = 100$ pF) moving radially outward on a trajectory offset by 10° from the antenna (see the illustration in the left panel).

The right panel in Fig. 2 shows the variation of the values of the geometric function with radial distance. $g_{SC}(r)$ is close to unity for the point charge located near the surface and then drops quickly with increasing distance. In the case investigated, this drop is steeper than $1/r$ because of the proximity of the antenna to the point charge since a significant fraction of the induced charge will occur on the antenna rather than the SC. The value of $g_{ANT}(r)$ increases quickly over a few mm distance as the point charge gets close to one end of the antenna. There is a noticeable drop in $g_{ANT}(r)$ at around radial distance $r = R_{SC} + 0.27 \cong 0.35$ m, where the antenna ends.

## 2.4   Electrostatic Model

The analytical model for a simplified case presented by *Shen et al.* [2021] calculated the time evolution of the charge on the SC in the form of:

$$\delta Q_{SC}(t) = Q_{SC,col} + \sum_{s=e,i} Q_{esc,s} \left( \frac{R_{SC}}{R_{SC}+v_s t} \right) + \int_0^t I_{dis}(\tau) d\tau. \tag{9}$$

The collected charge is the sum of collected electrons and cations, $Q_{SC,col} = Q_{e,col} + Q_{i,col}$, using Eqs. (2) and (3). The second term accounts for the induced charging from the escaping fractions of electrons and ions from the impact plasma. The simplifying assumptions used here are the spherical approximation of the SC. The escape of the electrons and ions can be effectively approximated as a point charge moving radially outward with their respective characteristic escape speeds ($v_s$, $s = e, i$). The last term in Eq. (9) accounts for the discharge current. The discharge current is due to the applied bias resistor for the case of the laboratory model and due to the charging environment for an SC operating in space. As the potential of the SC deviates from equilibrium due to charging from the dust impact plasma, the imbalance of the electron, ion, and photoelectron currents will drive the body back towards equilibrium [e.g., *Zaslavsky*, 2015]. Once the charging of the SC is calculated, the potential perturbation from the dust impact plasma is obtained as $\delta V_{SC}(t) = \delta Q_{SC}(t)/C_{SC,eff}$.

Here we generalize the simplified analytical model from *Shen et al.* [2021] by employing the capacitance matrix and geometric functions introduced above. The fundamental property of the model is that recollected and induced charges can be treated similarly when it comes to calculating the potential on the elements. The voltage perturbations can be calculated simultaneously in the following form:

$$
\begin{bmatrix} \delta V_{SC}(t) \\ \vdots \\ \delta V_{ANT,n}(t) \end{bmatrix} =
$$

$$
\begin{bmatrix} a_{00} & \dots & a_{0n} \\ \vdots & \ddots & \vdots \\ a_{n0} & \dots & a_{nn} \end{bmatrix} \begin{bmatrix} \delta Q_{SC,col} + \int_V \rho_{esc}(\vec{r},t)g_{SC}(\vec{r})\,dV + \int_0^t I_{SC,dis}(\tau)d\tau + \sum_{n=1}^n \int_0^t \frac{\delta V_{ANT,n}(\tau) - \delta V_{SC}(\tau)}{R_{base}}d\tau \\ \vdots \\ \delta Q_{ANT,col,n} + \int_V \rho_{esc}(\vec{r},t)g_{ANT,n}(\vec{r})\,dV + \int_0^t I_{ANT,dis,n}(\tau)d\tau - \int_0^t \frac{\delta V_{ANT,n}(\tau) - \delta V_{SC}(\tau)}{R_{base}}d\tau \end{bmatrix}.
$$

$$(10)$$

The elements of the charge vector on the right-hand side are similar to those presented in Eq. (9). The first term is the collected charge. The second term provides the induced charging from the escaping charges in form of a volume charge distributions, $\rho_{esc}(\vec{r},t) = \rho_{esc,e}(\vec{r},t) + \rho_{esc,i}(\vec{r},t)$, which accounts for both electrons and ions. This is the most general form that allows incorporating models of electrons and ion escaping, for example, by considering the initial velocity distributions and plume shapes of the respective species. The geometric functions in the integral enable considering the impact location with respect to the geometry of the SC and the antennas. The third term is the discharge from the environment. A fourth term is introduced, recognizing that the antennas and the spacecraft are electrically connected through a base resistance, $R_{base}$ [see, for example, *Bale et al.,* 2008]. This connection allows a current to flow and reduce the potential difference between the SC and the antennas. This last term may or may not be significant in comparison to the discharge through the environment, depending on the value of $R_{base}$ (typically on the order of $M\Omega$), the density of the ambient plasma, and the distance to the Sun, which would drive the photoelectron emission current.

# 3   Laboratory Setup

## 3.1   Model Spacecraft for Laboratory Studies

A new model SC has been developed and built based on the experience from the previous laboratory studies [*Nouzák et al.*, 2018, 2020; *Shen et al.*, 2021]. The new model SC is spherical with a radius $R_{SC} = 7.62$ cm and is equipped with four antennas that are arranged into one plane and spaced 90° apart (Fig. 3). Each antenna is a 27 cm long rod with a 1.6 mm diameter. The materials of both the SC and the antennas are stainless steel. The spherical SC is coated with graphite paint in order to provide it with a uniform surface potential [*Robertson et al.*, 2004]. The surfaces of the antennas were cleaned using organic solvents and are not coated. The circumference of the model SC is wrapped with a strip of tungsten (W) foil in the plane of the antennas. The foil is approximately 1.5 cm wide and is attached to the sphere by spot-welding. The purpose of the W foil is to provide the target surface for the dust impacts and make the measurements directly comparable to prior studies using the same material [*Nouzák et al.*, 2018, 2020; *Shen et al.*, 2021]. The surface of the foil was also cleaned using organic solvents.

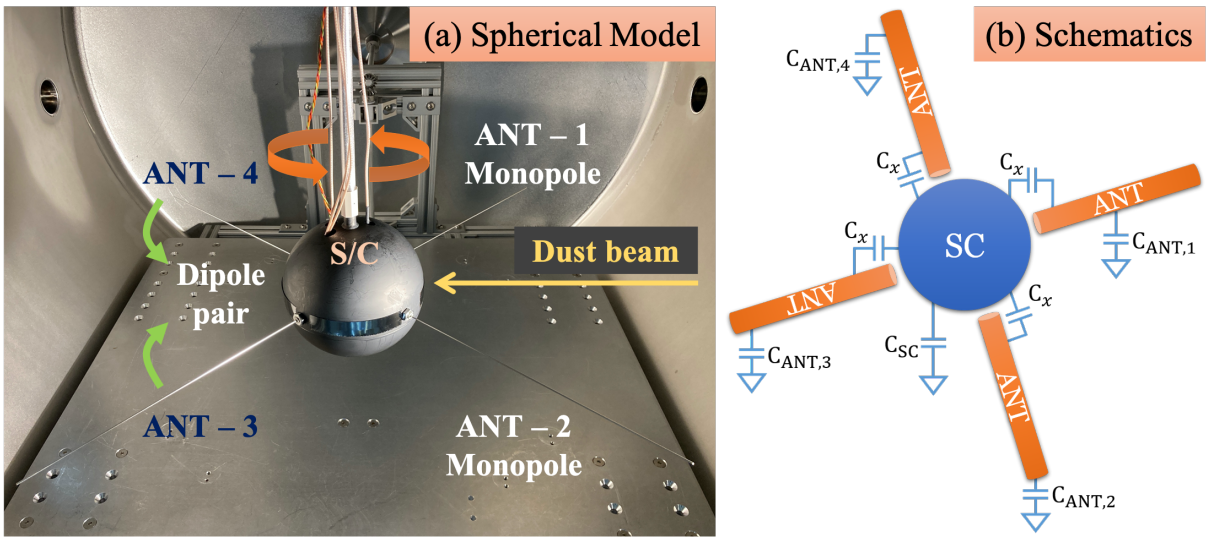

**Fig. 3:** *Left:* Photograph of the spherical model SC installed into the vacuum chamber. *Right:* Simplified schematics of the capacitances and the mutual coupling between the elements. See text for more detail.

Two of the antennas are configured as monopoles, and one pair operates in the dipole mode (Fig. 3). The sensing electronics is the same as described by *Shen et al.* [2021], and the boards are mounted within the hollow body of the model SC. Briefly, three instrumentation amplifiers are used to measure the voltage difference between the SC and two antennas, and between the dipole antenna pair. The bandwidth of the amplifiers is 270 Hz – 5 MHz, and they provide a voltage gain of 50. The circuits also allow the external biasing of the SC and the antennas. The same bias voltage is applied to each element (SC and four antennas) through individual biasing resistors. The effective values of these resistors are $R_{BIAS,A} = 5\ M\Omega$ for each of the antennas, and $R_{BIAS,SC} =$ 2.5 $M\Omega$ for the SC. The bias resistors are combined with the effective capacitances of the antennas and the SC and provide the characteristic time constants for each of the elements, $R_{BIAS,A}C_{A,eff}$ for each of the antennas, and $R_{BIAS,SC}C_{SC,eff}$ for the SC body. The measured waveforms are recorded using a fast-digitizing oscilloscope.

The model SC is mounted onto a vertical shaft such that the antennas are in the horizontal plane (Fig. 3). The shaft is connected to a rotary feedthrough that positions the model SC into the center of a large vacuum chamber. The vacuum chamber is 1.2 m in diameter, 1.5 m long, and is evacuated to $\sim 10^{-6}$ Torr using oil-free vacuum pumps.

The measurements described below were performed using the Iron (Fe) dust sample that also makes them comparable to prior studies conducted by *Collette et al.* [2015; 2016], *Nouzák et al.* [2018; 2020], and *Shen et al.* [2021]. The accelerator facility used for the studies is described by *Shu et al.* [2012]. The mass and velocity for each accelerated particle are calculated from pick-up tube detector signals provided by the facility. A comprehensive measurement campaign has been conducted in order to investigate the effects of impact velocity, impact location, bias voltage, and antenna operation mode on the recorded waveforms. The focus of this article is limited to impact the impact velocity range of 20 – 40 km/s in order to evaluate the validity of the proposed signal generation mode. The dust beam is pointed at the center of the spherical model SC, and thus the impacts occur normal to the W target foil.

## 3.2   Capacitance Matrix

The elastance matrix is a key element in implementing the electrostatic antenna signal generation model presented in Sec. 2. This section describes how to calculate this matrix for the model SC described above, how to calculate it for an actual SC, and the physical meaning of the elements. Appendix A provides the details for an SC with a single antenna.

The elastance matrix $[\boldsymbol{a}]$ is the inverse of the capacitance matrix, $[\boldsymbol{b}] = [\boldsymbol{a}]^{-1}$. The dimension of the matrices is $(n + 1) \times (n + 1)$, where $n$ is the number of independent antennas in the system, and the +1 refers to the SC body. A diagonal element $b_{ii}$ in $[\boldsymbol{b}]$ is called the self-capacitance and represents the capacitance of the $i$-th object in a configuration, where all other elements of the system are grounded [*Jackson*, 1999]. For example, assuming that $i = 0$ refers to the SC, and considering the schematics shown in Fig. 3, it follows that $b_{00} = C_{SC} + 4C_X$. Here $C_{SC}$ is the physical capacitance of the SC body only (without the antennas), and an assumption has been made that the mutual capacitances between the SC and each of the antennas are equal. The non-diagonal elements $b_{ij}$ represent the negative value of the mutual capacitances between elements $i$ and $j$. For example, $b_{01} = -C_x$, which is the mutual capacitance between the SC and antenna #1. The matrix is symmetric, i.e., $b_{ij} = b_{ji}$ The capacitance matrix for the model SC used in the laboratory measurements can be written as:

$$[b] \cong \begin{bmatrix} C_{SC} + 4C_x & -C_x & -C_x & -C_x & -C_x \\ -C_x & C_{ANT,1} + C_x & 0 & 0 & 0 \\ -C_x & 0 & C_{ANT,2} + C_x & 0 & 0 \\ -C_x & 0 & 0 & C_{ANT,3} + C_x & 0 \\ -C_x & 0 & 0 & 0 & C_{ANT,4} + C_x \end{bmatrix} \cong$$

$$\begin{bmatrix} 52 & -6.5 & -6.5 & -6.5 & -6.5 \\ -6.5 & 16 & 0 & 0 & 0 \\ -6.5 & 0 & 17 & 0 & 0 \\ -6.5 & 0 & 0 & 19 & 0 \\ -6.5 & 0 & 0 & 0 & 16.5 \end{bmatrix} \text{pF} \tag{11}$$

The last term in Eq. (11) presents the values determined from measurements made on the model SC. It makes the simplifying assumption that the antenna-to-antenna mutual capacitances are negligible, and the validity of this assumption is provided below. The details of the measurements and calculations are presented in Appendix B. The physical capacitances of the elements of the system are $C_{SC} = 26$ pF, $C_{ANT,1} = 9.5$ pF, pF, $C_{ANT,2} = 10.5$ pF, $C_{ANT,3} = 12.5$ pF, and $C_{ANT,4} = 10$ pF. These values are significantly higher than the capacitances of the objects based on their physical dimensions alone. There are several reasons for this. The model spacecraft is installed into a vacuum chamber with grounded walls. Further, the electronic boards mounted within the hollow spherical model SC are referenced to ground, which increases the capacitances of the SC, but also those of the antennas. In other words, for the case of the laboratory model SC, the parasitic capacitances $C_{WP}$ from the cabling and preamplifier (see Sec. 2.2) contribute to the antennas' capacitances with respect to ground, rather than to the mutual capacitance towards the SC.

The capacitance matrix can also be calculated using standard numerical electrostatic solvers. The simulated matrix for the model SC is the following:

$$[\boldsymbol{b}]^{sim} = \begin{bmatrix} +11.800 & -1.2100 & -1.2100 & -1.2100 & -1.2100 \\ -1.2100 & +3.1300 & -0.0728 & -0.0267 & -0.0728 \\ -1.2100 & -0.0728 & +3.1300 & -0.0728 & -0.0267 \\ -1.2100 & -0.0267 & -0.0728 & +3.1300 & -0.0728 \\ -1.2100 & -0.0728 & -0.0267 & -0.0728 & +3.1300 \end{bmatrix} \text{pF}. \tag{12}$$

The elements of $[\boldsymbol{b}]_{sim}$ are considerably different from those of $[\boldsymbol{b}]$ from Eq. (11). This is because the simulation result can only include the geometry of the system. For example, the simulated physical capacitance of the first antenna is only $C_{ANT,1} = \sum_{j=0}^{4} b_{1,j} \cong 1.75$ pF, which is significantly lower than the ~10 pF capacitance as determined from the actual measurements. The difference is due to the contribution from cabling, electronics, etc. Similarly, the geometric contribution to the mutual capacitance between the SC and the antenna is only $C_G = 1.21$ pF. The results in Eq. (12) also demonstrate that the mutual capacitance between two adjacent antennas (~0.073 pF) is much smaller than that of the SC-to-antenna mutual capacitance and thus can be neglected for the model SC system, where the antennas are placed far from one another. On the other hand, the antennas on the Cassini or STEREO missions are mounted on the base and thus their mutual capacitance will likely be significant [*Gurnett et al.*, 2004; *Bale et al.*, 2008].

## 3.3   Constructing the Capacitance Matrix for a Spacecraft

This article aims to provide the framework for the improved analysis of the dust impact signals measured by antenna instruments on space missions. This section provides the recommendation on how to estimate the capacitance matrix for any spacecraft with adequate accuracy. The first step is importing a reasonably detailed Computer-Aided Design (CAD) model of the SC with the antennas into the software tool and calculate the simulated capacitance matrix. This matrix can be deconstructed following the definitions provided in Eq. (11) in order to determine the values of $C_{SC}$, the capacitances for each of the antennas, and the mutual capacitances between the elements. Alternatively, the antenna capacitances published by the instrument team may also be used. The mutual capacitances between the antennas may or may not be negligible, depending on their dimensions and arrangement. At this point, the mutual capacitance values include only the contributions from the geometry of the system ($C_G$). The contribution from the base capacitance ($C_B$) and the combined effect of cables and preamp input capacitance ($C_{WP}$) are typically provided by the instrument team. The combined value is called the stray capacitance, $C_{stray} = C_B + C_{WP}$ [e.g., *Bale et al.,* 2008]. The mutual capacitance between the antennas and the SC is then given as $C_x = C_G + C_{stray}$. The stray capacitance may already include the effect of $C_G$, and in this case, it is simple $C_x = C_{stray}$. The capacitance matrix defined by Eq. (11) can from here be reconstructed from the determined values.

## 3.4   Induced Charging for the Model SC

Numerical solvers can provide the induced potential from a test point charge as described in Sec. 2.3. This section presents the calculated induced potentials for the three geometries investigated in the laboratory experiments, namely with the dust impact occurring in between antennas #1 and #2 at 10°, 30°, and 45° measured from antenna #1 (see Fig. 3). Figure 4 shows the induced potentials on the elements for these three configurations. The point test charge has the value of $Q_{test} = 100$ pC and is moving on a radial trajectory, starting 0.3 mm from the surface of the SC. This setup is similar to that shown in Fig. 2 for a simplified case, and the relevant capacitance matrix is $[b]^{sim}$ from Eq. (12). The corresponding simulated elastance matrix is $[a^{sim}] = [b^{sim}]^{-1}$.

When the test charge is closest to the SC body, the induced charge on the antennas is negligible, and the SC potential is $\delta V_{SC} = a_{0,0}^{sim} \, Q_{test} = 10.18$ V, following Eq. (6). The meaning of the first element of the elastance matrix is the inverse of the effective capacitance of the SC. For the same conditions, the potentials on all of the antennas are the same. For example, on antenna #1 the potential is $\delta V_{ANT,1} = a_{1,0}^{sim} \, Q_{test} = 4.17$ V, and this relation defined the physical meaning of the corresponding element of $[a^{sim}]$. The mutual capacitive coupling from the SC to the antenna is significant and will affect the measured voltage ($\delta V_{meas} = \delta V_{ANT,1} - \delta V_{SC}$).

The SC potential for all three cases decreases monotonically as the test charge is moving radially outward. For the 10° case, the induced potential on antenna #1 is increasing sharply due to the test charging getting in the close vicinity of this element. At its maximum of about $\delta V_{ANT,1} = 12.5$ V, the induced charge on this antenna is about 37% of $Q_{test}$, as it can be calculated from the $0.37 \, a_{1,1}^{sim} \, Q_{test} \cong 12.5$ V relation. Antenna #1 has a pronounced, albeit lower maximum potential value for the 30° case as well. The values of the relevant coefficients from the elastance matrix are: $a_{0,0}^{sim} = 1.018 \times 10^{11} \text{F}^{-1}$, $a_{1,0}^{sim} = 0.4166 \times 10^{11} \text{F}^{-1}$, and $a_{1,1}^{sim} = 3.369 \times 10^{11} \text{F}^{-1}$.

The potential profiles on the antennas on the far side of the SC (#3 and #4) mostly follow that of the mutual coupling from the SC, with only a small contribution from direct induced charging from the point charge. The dipole signal between these antennas ($\delta V_{dipole} = \delta V_{ANT,3} - \delta V_{ANT,4}$) is small, albeit zero only for the symmetric 45° case. Were antennas #1 and #2 connected as a dipole, the measured signal ($\delta V_{ANT,1} - \delta V_{ANT,2}$) would be significant and a strong function of the impact location and the expansion of the plasma plumes with respect to the antenna geometry.

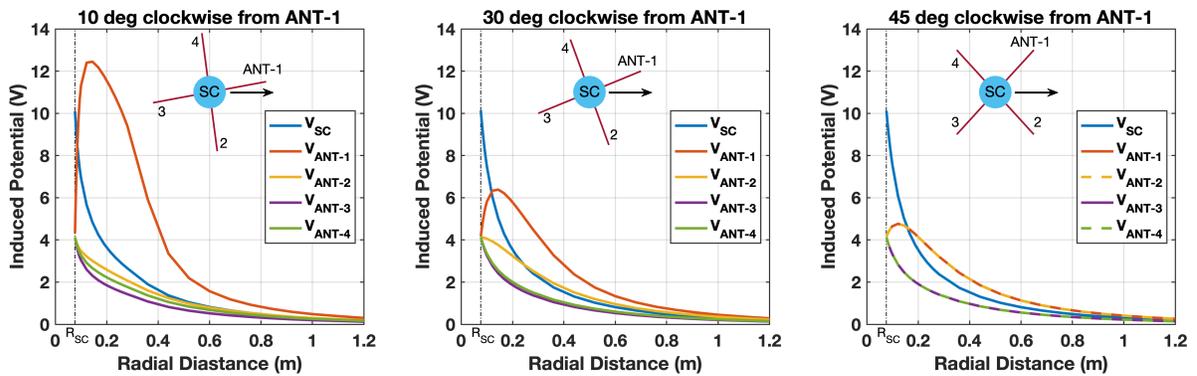

**Fig. 4:** Induced potentials on the SC and the four antennas from a test charge moving on a radial trajectory for three different configurations relative to antenna #1 (10°, 30°, and 45°). The potentials are calculated numerically for a $Q_{test} = 100$ pF test charge and the geometry of the model SC used in the experiments. See text for details.

# 4  Data Analysis

This article presents a set of the collected data with a goal to demonstrate the validity of the proposed electrostatic model. The main reason for this limitation is the incomplete understanding of the properties of the expanding impact plasma cloud. The model assumes that the expanding electrons have an isotropic velocity distribution; however, it is not obvious what the shape of the expanding plume of ions is. Section 3.4 above demonstrates that the voltages induced on the antennas are rather sensitive to how close the escaping charges get to the antennas. In other words, an ion plume in a shape of a narrow pencil beam would generate different induced voltages than an ion plume with a wider, conical shape, for example. In order to avoid the confusion between competing mechanisms and geometrical effects, a subset of the measurements was collected with antenna #2 connected to ground potential ($V_{ANT,2} = 0$). This means that the corresponding channel measures directly the inverted potential of the SC, i.e., $\delta V_{meas} = 0 - \delta V_{SC}$. Accounting for the duration and shape of the ion plume goes beyond the current capabilities of the signal fitting routine described below. Instead, we follow the simplification from *Shen et al.* [2021], and the escaping electrons and cations are modeled as point charges moving radially away with their respective escape velocities. The detailed analysis of the data is left as a task for the future.

## 4.1   Fitting Routine

The numerical and simplified version of Eq. (10) is used to fit the laboratory measurements to validate the electrostatic model and calculate some of the fundamental parameters of the escaping plasma plume. Since antenna #2 is grounded, there are only four active elements in the experimental system. The capacitance matrix with ($4 \times 4$) elements is constructed following the description provided in Sec. 3.2, i.e.,

$$[\boldsymbol{b^{exp}}] \cong \begin{bmatrix} 52 & -6.5 & -6.5 & -6.5 \\ -6.5 & 16 & 0 & 0 \\ -6.5 & 0 & 19 & 0 \\ -6.5 & 0 & 0 & 16.5 \end{bmatrix} \text{pF},\tag{13}$$

and the elastance matrix is calculated as $[\boldsymbol{a^{exp}}] = [\boldsymbol{b^{exp}}]^{-1}$. The geometric functions were calculated for radial trajectories for impact locations $10°$, $30°$, and $45°$ offset from antenna #1 (see Fig. 4). The form of Eq. (10) for numerical calculations is:

$$\begin{bmatrix} \delta V_{SC}(i) \\ \delta V_{ANT,1}(i) \\ \delta V_{ANT,3}(i) \\ \delta V_{ANT,4}(i) \end{bmatrix} = [\boldsymbol{a^{exp}}] \begin{bmatrix} \delta Q_{SC,col} + \left[ Q_{esc,e}g_{SC}(\vec{r_e}(i)) + Q_{esc,i}g_{SC}(\vec{r_i}(i)) \right] - \sum_{k=0}^{i-1} \frac{\delta V_{SC}(k)}{R_{base,SC}}\Delta t \\ \delta Q_{ANT,1,col} + \left[ \zeta_e Q_{esc,e}g_{ANT,1}(\vec{r_e}(i)) + \zeta_i Q_{esc,i}g_{ANT,1}(\vec{r_i}(i)) \right] - \sum_{k=0}^{i-1} \frac{\delta V_{ANT,1}(k)}{R_{base,A}}\Delta t \\ \delta Q_{ANT,3,col} + \left[ Q_{esc,e}g_{ANT,3}(\vec{r_e}(i)) + Q_{esc,i}g_{ANT,3}(\vec{r_i}(i)) \right] - \sum_{k=0}^{i-1} \frac{\delta V_{ANT,3}(k)}{R_{base,A}}\Delta t \\ \delta Q_{ANT,4,col} + \left[ Q_{esc,e}g_{ANT,4}(\vec{r_e}(i)) + Q_{esc,i}g_{ANT,4}(\vec{r_i}(i)) \right] - \sum_{k=0}^{i-1} \frac{\delta V_{ANT,4}(k)}{R_{base,A}}\Delta t \end{bmatrix}.$$

$$\tag{14}$$

The numerical calculations are performed over discrete time steps $\Delta t$, and index $(i)$ represents time $t(i) = i \times \Delta t$, where $t = 0$ marks to the instance of the impact. The dust impact plasma at this point is approximately the size of the impacting particle with net zero charge, and thus the initial conditions are:

$$\begin{bmatrix} \delta V_{SC}(0) \\ \delta V_{ANT,1}(0) \\ \delta V_{ANT,3}(0) \\ \delta V_{ANT,4}(0) \end{bmatrix} = \begin{bmatrix} 0 \\ 0 \\ 0 \\ 0 \end{bmatrix} V \qquad (15)$$

The term on the right-hand side of Eq. (14) represents the time evolution of the charge balance on each of the elements, including both the collected and induced charges. The first term in the vector is the collected charge given by Eq. (2) or (3) and is a constant. The second term is the sum of the induced charges from the escaping electrons and ions. These charges are weighted by the geometric function, where position vectors of the assumed point charges evolve over time. The position vectors are $\vec{r_e}(i) = \vec{r}_{imp} + \hat{r}\, v_e\, i\Delta t$ for the escaping electrons, and $\vec{r_i}(i) = \vec{r}_{imp} + \hat{r}\, v_i\, i\Delta t$ for the escaping ions. Here $\vec{r}_{imp}$ is the location of the impact, $\hat{r}$ is the radial unit vector at the location of the impact, and $v_e$ and $v_i$ are the electron and ion escape speeds, respectively. Two new parameters, $\zeta_e$ and $\zeta_i$, are introduced for the induced charge term for antenna #1, which is the closest to the dust impact locations. These function as free fitting parameters that allow accounting for the deviations of the shape of expanding plasma plume from a point charge moving radially outward with a constant speed.

The last term is the integral (or summation in the numerical form) of the discharge current. This term is somewhat different from that presented in Eq. (10) as each element in the model SC can only discharge through its individual bias resistor, which is referenced to ground. This discharge current will drive the voltage perturbations on each of the elements to zero for $t \rightarrow \infty$. Equation (14) can be easily modified to be applicable for SC operating in the space environment. This requires implementing the discharge current from the ambient plasma (Sec. 2.4) and modifying the last term following Eq. (10).

Solving Eq. (14) provides the time evolution of the physical voltages on each of the elements. This equation, however, does not include the effects of the electronics, namely its gain and limited bandwidth. The actual waveforms measured by the antenna instrument can be calculated by convolving the physical voltages by the impulse response of the electronics. Figure 5 shows the impulse response of the electronics used in the model SC that was calculated using industry standard SPICE (Simulation Program with Integrated Circuit Emphasis) software from the schematics of the electronics and the parts used.

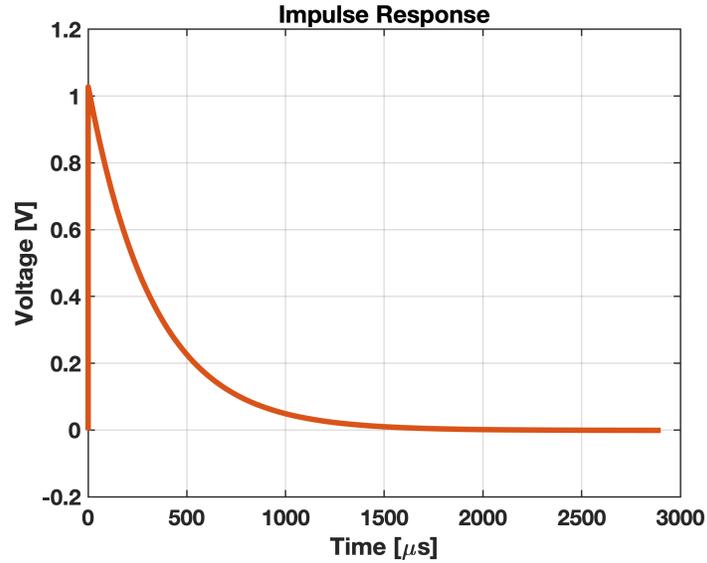

**Fig. 5:** Impulse response of the electronics used in the model SC used for the laboratory measurements.

## 4.2   Impacts at 45°

Figure 6 shows a typical set of antenna signals measured for an impact location in between antennas #1 and #2, i.e., at 45° off from antenna #1 (Fig. 3). The bias voltage on all elements is set to 0V. The two measured signals shown are $\delta V_{mono,1} = \delta V_{ANT,1} - \delta V_{SC}$, and $\delta V_{mono,2} = -\delta V_{SC}$, since $\delta V_{ANT,2} = 0$. The first thing to notice is that the two signals are significantly different, with the maximum on $\delta V_{mono,1}$ occurring earlier in time. This indicates that induced charging from the escaping electrons contributes significantly to the waveforms measured by antenna #1. Once the escaping electrons and ions expanded beyond the length of the antenna, signal $\delta V_{mono,1}$ relaxes back to sensing only the charge that has been collected on the SC. This means $\delta V_{mono,1} \cong (a_{1,0}^{exp} - a_{0,0}^{exp})(\delta Q_{i,col} - \delta Q_{e,col})$, where the last term is the net collected charge. The values of the relevant elastance matrix are $a_{0,0}^{exp} = 2.24 \times 10^{10} \text{F}^{-1}$ and $a_{1,0}^{exp} = 0.91 \times 10^{10} \text{F}^{-1}$, which provide the estimate for the ratio of $\delta V_{mono,1}/\delta V_{mono,2} = (a_{0,0}^{exp} - a_{1,0}^{exp})/a_{0,0}^{exp} \cong 0.6$. The measured ratio is somewhat larger for the particular example shown; nevertheless, the model explains why the signal measured by antenna #1 drops below the $\delta V_{mono,2}$ signal at around $t = 30$ μs. It will be show below that this is roughly the time for the cations to expand beyond the length of the antennas. The feature of signal $\delta V_{mono,1}$ crossing and dropping below signal $\delta V_{mono,2}$ is typical for all measurements taken in this configuration.

The start of the waveforms is similar to those observed by *Nouzák et al.* [2018] or *Shen et al.* [2021]. Briefly, the sharp negative drop is known as the preshoot, and is due to the fast-escaping electrons that leave a net positive charge near the SC, which temporarily drives the SC potential to $\delta V_{SC} > 0$. After reaching a minimum, the waveform signals increase due to the escape of the slower cations. The rate of the increase, however, is different for the two signals. This is because antenna #1 also senses the induced charges from the cloud of escaping cations, as indicated in Fig. 4. This effect also drives $\delta V_{mono,1}$ to be more positive than $\delta V_{mono,2}$ for the duration of the cation expansion over the length of the antennas. Once the escape of the electrons and cations is complete, the SC is left with a net negative charge. This is due to the difference in the properties of electron and cation clouds emerging from the impact plasma. While the cations are expanding in the form of a plume that moves away from the impact location, the electrons have an isotropic distribution. The latter results in the recollection of about half of the electrons by the SC for the investigated case of no bias potential applied on the elements. The charge collected on the SC discharges through the bias resistor with a characteristic time constant, as described in Sec. 3.1.

The model described above allows for fitting the waveforms and determining some of the key parameters of the impact plasma. For the data shown in Fig. 6, these parameters are $Q_{IMP} = 1.13 \times 10^{-13}$ C, $\kappa = 0.44$, and $v_i = 12.0$ km/s. The latter two values are in good agreement with prior measurements, and electron expansion speed is set to $v_e \cong 10^3$ km/s [*Shen et al.,* 2021]. The values of the two new parameters introduced in Eq. (14) are $\zeta_e = 2.50$ and $\zeta_i = 1.24$. It would not be possible to properly match the negative and positive amplitudes of the $\delta V_{mono,1}$ waveform without these auxiliary parameters. The current numerical model allows treating the escaping electrons and cations as point charges only and are moving radially outward with constant velocities. This physical picture, however, is oversimplified. The electrons escaping from the plasma cloud have an isotropic distribution, and thus some fraction of the electrons will get in the close vicinity of the antenna. This results in a stronger induced charging signal than the model with the current geometric functions can reproduce. Similarly, parameter $\zeta_i > 1$ indicates that the cations also get closer to the antenna than they would from moving on a radial line. In other words, the fitting results indicate that the cations are expanding in form of a divergent plume rather than as a narrow pencil beam. The more detailed analysis of the data will require (1) having a solution for the geometric function fall all locations in the vicinity of the SC, and (2) employing realistic models for the expansion of the electrons and ions that would also allow calculating their charge densities as a function of time and location as presented in Eq. (10).

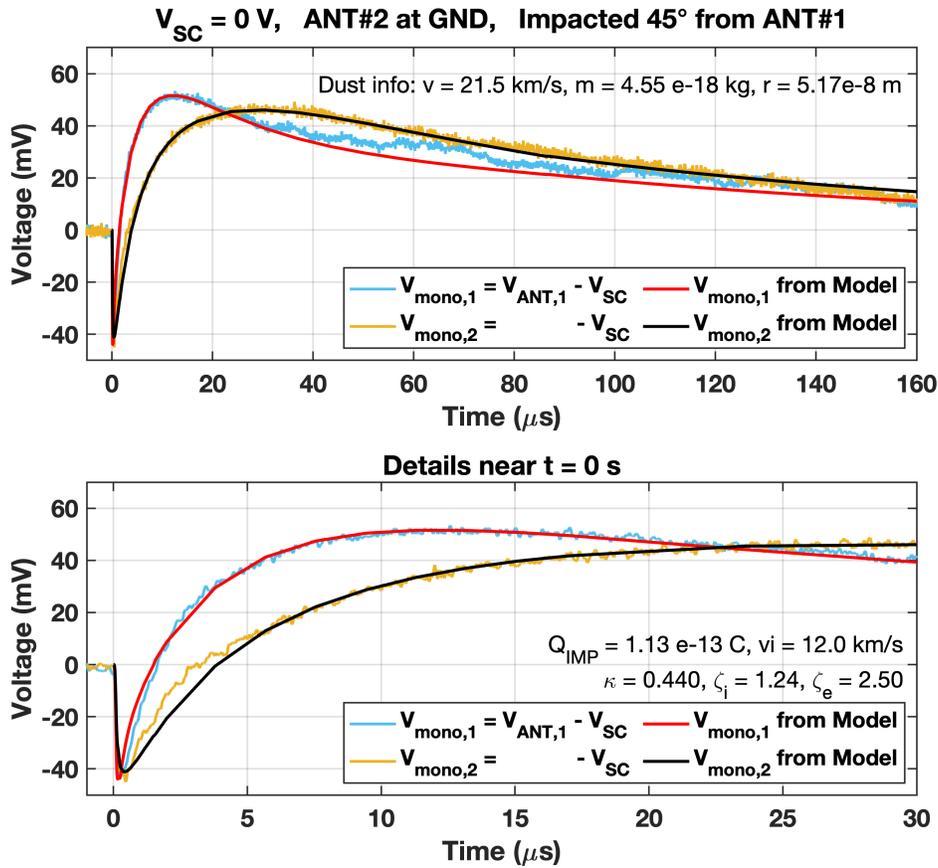

**Fig. 6:** Typical waveforms measured in the laboratory for an impact location at 45° from antenna #1. The properties of the impacting dust particle are provided in the top panel. The bottom panel provides the details in the early phases of the impact plasma expansion. See text for more details.

## 4.3   Impacts at 10°

Figure 7 shows a set of typical waveforms for a dust impact location 10° offset from antenna #1. Many features are similar to the 45° case treated above. The obvious difference is the much more pronounced contribution from the induced charge signal on antenna #1, which is closest to the impact location, and so do the expanding plasma cloud. In addition, the model does not provide as good of an agreement with the data as in the 45° impact location. The impact location is in the close proximity of the antenna base (separated by about $1 - 1.5$ cm), and the diverging cation plume results in relatively large differences between the measurements and the simplified expansion model. Cations in a conical expansion plume would get close to the antenna faster than in the case of a radial pencil beam that is moving close to parallel to the antenna. As a result, there is an ambiguity in determining the ion expansion speed from the fit. The 'best' fit to the data in Fig. 7 was guided by approximately matching the intersect point between the $\delta V_{mono,1}$ and $\delta V_{mono,2}$ signals at around $t = 30$ μs.

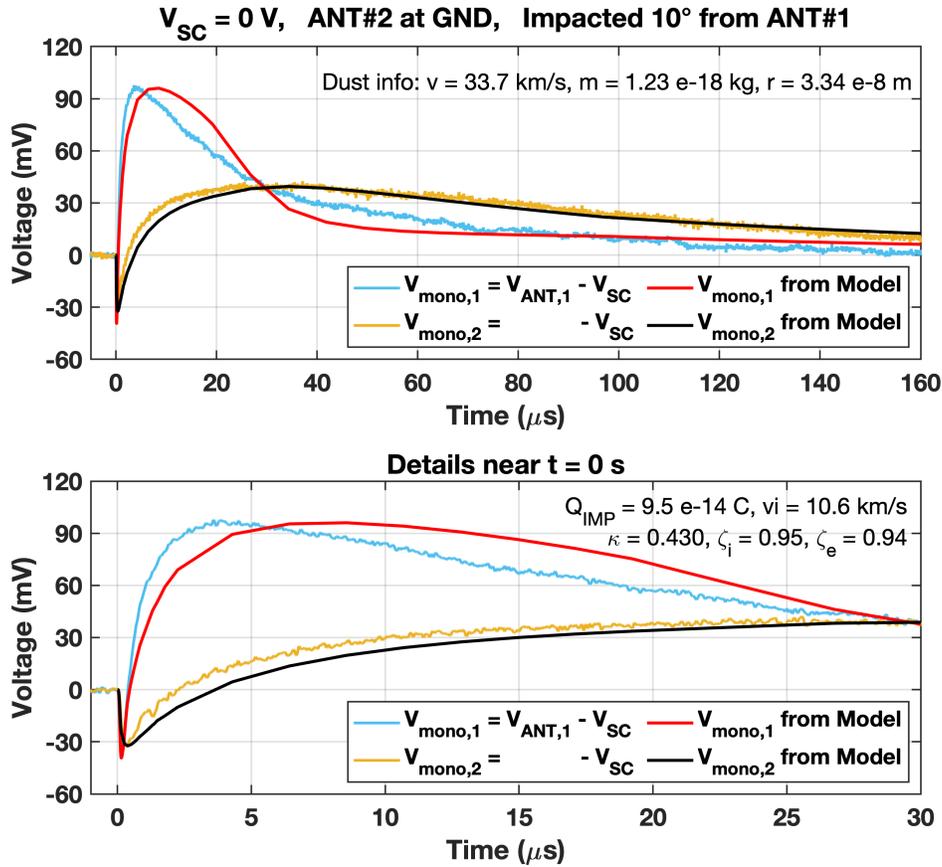

**Fig. 7:** Typical waveforms measured in the laboratory for an impact location at 10° from antenna #1. The properties of the impacting dust particle are provided in the top panel. The bottom panel provides the details in the early phases of the impact plasma expansion. See text for more details.

The fitting parameters determined from the model are $Q_{IMP} = 9.5 \times 10^{-14} C$, $v_i = 10.6$ km/s, and $\kappa = 0.43$, with the latter two in good agreement with the results for the 45° impact location. Both zeta parameters are to unity, $\zeta_e = 0.94$ and $\zeta_i = 0.95$, meaning that the amplitudes of the waveforms are reproduced accurately by the model. Generally, the model provides a good match for the entire $\delta V_{mono,2}$ waveform and the beginning and the end of the $\delta V_{mono,1}$ waveform.

## 5   Summary and Conclusions

The article presents the general electrostatic model for understanding the generation of the transient voltage perturbations detected by antenna instruments. The matrix form provides a convenient way to track the interaction between the elements and calculate the voltage differences in between. In addition, the elastance matrix offers a straightforward course of calculating the effective capacitances of the elements needed to convert the measured voltages to charge appropriately, or vice versa. Overall, the presented model will improve data analysis fidelity and calculate the impact charge from the dust particle, which in turn allows determining its mass. This is, of course, under the assumption that we know the impact speed, SC potential, and the effective temperatures of the electrons and cations of the impact plasma.

It is remarkable how well the model reproduces the measured waveforms, using only a small set of fitting parameters. This fact confirms the suggestions of prior studies that there are two primary signal generation mechanisms: one due to the recollected charge from the impact plasma and the second from the induced charge from the escaping fraction of the impact plasma. One of the fundamentals of the model is the recognition that the collected and induced charges can be treated similarly. If desired, the model can be easily augmented to include the charge collection by the antennas for even higher fidelity. This may be significant for dust impacts occurring in the close vicinity of an antenna base.

*Shen et al.* [2021] presented a simplified model applicable to the simplified case, where the antenna is far from the impact location. The full model presented in this article employs the geometric functions to account for the generation of induced charge signals on the antennas. The measurements have shown that the induced charge is significant even for impacts relatively far from the antenna base. This has several important consequences: (1) The model can be used to analyze the wide variety of expected waveforms from the dust impact signals as a function of impact location (and other parameters, e.g., those of the ambient plasma). Such analysis would be useful for recognizing valid dust impact events. (2) There is a promising outlook that the detailed analysis of the waveforms detected by multiple antennas can be used to constrain the impact location on the SC body, which in turn could provide useful information on the orbital elements of the impacting particle. The induced charge signal from a plasma plume is unique for each antenna and impact location. The small variations between antenna waveforms could thus reveal the origin of dust particles. (3) The previous point can be turned around, and antenna waveforms for a known dust impact location can be used to characterize the properties of dust impact plasma plumes. Our understanding of dust impact plasmas is surprisingly limited, and antennas may provide an elegant way to learn about the expansion characteristics of the electrons and cations. This method would be applicable both for laboratory measurements and data collected by space missions. (4) It may be worth revisiting the efficiency of dust impact detection for antennas operating in dipole mode. The presented model could be employed to analyze the variety of impact waveforms expected in this mode, which are significantly different from those measured in the monopole mode.

# Appendix A – Spacecraft with a Single Antenna

This Appendix presents illustrative exercise calculations for the simple case of an SC with one antenna that demonstrates: (1) that the capacitance matrix is properly tracking the effective capacitance of the system, and (2) the physical meaning of the elastance matrix. The '$\delta$' is omitted in these equations for simplicity. The potential of elements of the simple SC – antenna system is written using the elastance matrix $[a]$:

$$\begin{bmatrix} V_{SC} \\ V_{ANT} \end{bmatrix} = \begin{bmatrix} a_{00} & a_{01} \\ a_{10} & a_{11} \end{bmatrix} \begin{bmatrix} Q_{SC,col} \\ Q_{ANT,col} \end{bmatrix}. \tag{A.1}$$

The charge can be calculated from the voltage using the inverse form,

$$\begin{bmatrix} Q_{SC} \\ Q_{ANT} \end{bmatrix} = \begin{bmatrix} b_{00} & b_{01} \\ b_{10} & b_{11} \end{bmatrix} \begin{bmatrix} V_{SC} \\ V_{ANT} \end{bmatrix} = \begin{bmatrix} C_{SC} + C_x & -C_x \\ -C_x & C_{ANT} + C_x \end{bmatrix} \begin{bmatrix} V_{SC} \\ V_{ANT} \end{bmatrix}, \tag{A.2}$$

where $[b] = [a]^{-1}$ is the capacitance matrix. The right-hand side of the equation above demonstrates the physical meaning of the elements (see also Sec. 3.2 and Eq. (13)). The charges on the SC and the antenna can be expressed as:

$$Q_{SC} = (b_{00} + b_{01})V_{SC} + b_{01}(V_{ANT} - V_{SC})$$
$$Q_{ANT} = b_{10}(V_{SC} - V_{ANT}) + (b_{11} + b_{10})V_{ANT}, \tag{A.3}$$

where the potential differences between the SC and antenna are introduced. By substituting for the physical meaning of the matrix elements from Eq. (A.2) we obtain:

$$Q_{SC} = C_{SC}V_{SC} + C_x(V_{SC} - V_{ANT})$$
$$Q_{ANT} = C_x(V_{ANT} - V_{SC}) + C_{ANT}V_{ANT}, \tag{A.4}$$

This set of equations can be solved for the case where a charge is deposited only onto the SC, i.e., $Q_{ANT} = 0$. Expressing $V_{ANT} = V_{SC} C_x/(C_x + C_{ANT})$ from the second equation, and substituting to the first, we obtain

$$Q_{SC} = C_{SC}V_{SC} + C_x\left(V_{SC} - \frac{c_x}{c_x + c_{ANT}}V_{SC}\right). \tag{A.5}$$

And from here the effective capacitance is obtained as

$$\frac{Q_{SC}}{V_{SC}} = C_{eff,SC} = C_{SC} + \frac{c_x c_{ANT}}{c_x + c_{ANT}}. \tag{A.6}$$

The effective capacitance of the antenna can be calculated similarly.

The last thing to demonstrate is the physical meaning of the diagonal terms of the elastance matrix. Following the rules of matrix transformation, we obtain from Eq. (A.2)

$$[\boldsymbol{a}] = \frac{1}{(C_{SC}+C_x)(C_{ANT}+C_x)-C_x^2}\begin{bmatrix} C_{ANT}+C_x & C_x \\ C_x & C_{SC}+C_x \end{bmatrix}. \tag{A.7}$$

It can be shown easily from here that $a_{00}^{-1} = C_{eff,SC}$, and $a_{11}^{-1} = C_{eff,ANT}$. In other words, the diagonal elements conveniently provide effective capacitances for each element of the system.

# Appendix B – Capacitance Measurements

This appendix provides the details of how the elements of the Maxwell capacitance matrix presented in Eq. (11) were determined for the model SC. The measurements described below were performed with the model SC installed into the vacuum chamber, i.e., in the same conditions as dust impact measurements were made. The measurements were performed using a function generator and a $C_{test} = 10$ pF test capacitor. The function generator was configured to output a square wave with a $\Delta V = 50$ mV amplitude, and this signal was applied onto the SC or antenna elements through the test capacitor. The magnitude of the injected test charge was thus $Q_{test} = \Delta V C_{test} = 0.5$ pC. The response of the model SC's electronics to the test charge input was recorded using a fast-digitizing scope. The output signals then were fitting utilizing an industry standard SPICE (Simulation Program with Integrated Circuit Emphasis) software tool to calculate the net capacitance sensed on the input. Figure A1 illustrates one instance of the waveform generator output and the recorded signal in response to the injected test charge.

The first set of measurements were performed following the definition of the diagonal elements of the Maxwell capacitance matrix. In these measurements, all but one element of the system was grounded in order the calculate the net capacitance. The list below provides the results of these measurements:

SC:              $C_{test} + C_{SC} + 4C_x \cong 62 pF$

Antenna 1:     $C_{test} + C_{ANT,1} + C_x \cong 26.0 pF$

Antenna 2:     $C_{test} + C_{ANT,2} + C_x \cong 27.0 pF$

Antenna 3:     $C_{test} + C_{ANT,3} + C_x \cong 29.0 pF$

Antenna 4:     $C_{test} + C_{ANT,4} + C_x \cong 26.5 pF$ .

This set of equations already include the simplifying assumption that the mutual capacitances between the SC and the antennas are the same. Since there are six unknowns and only five equations, further measurements were necessary. Additional measurements were made, where the test charge was injected into one of the antennas while keeping the remaining three antennas grounded but allowing the SC to float. The output signals of these measurements were recorded as well. And as a final step, in a numerical procedure, the value of $C_x$ was varied between $4 - 8$ pF with steps of 0.5 pF in order to find the solution that provided the best fit for all measurements. This exercise resulted in $C_x = 6.5$ pF as the best value, which then provides the following capacitances: $C_{SC} = 26\ pF$, $C_{ANT,1} = 9.5\ pF$, $C_{ANT,2} = 10.5\ pF$, $C_{ANT,3} = 12.5\ pF$, and $C_{ANT,4} = 10\ pF$.

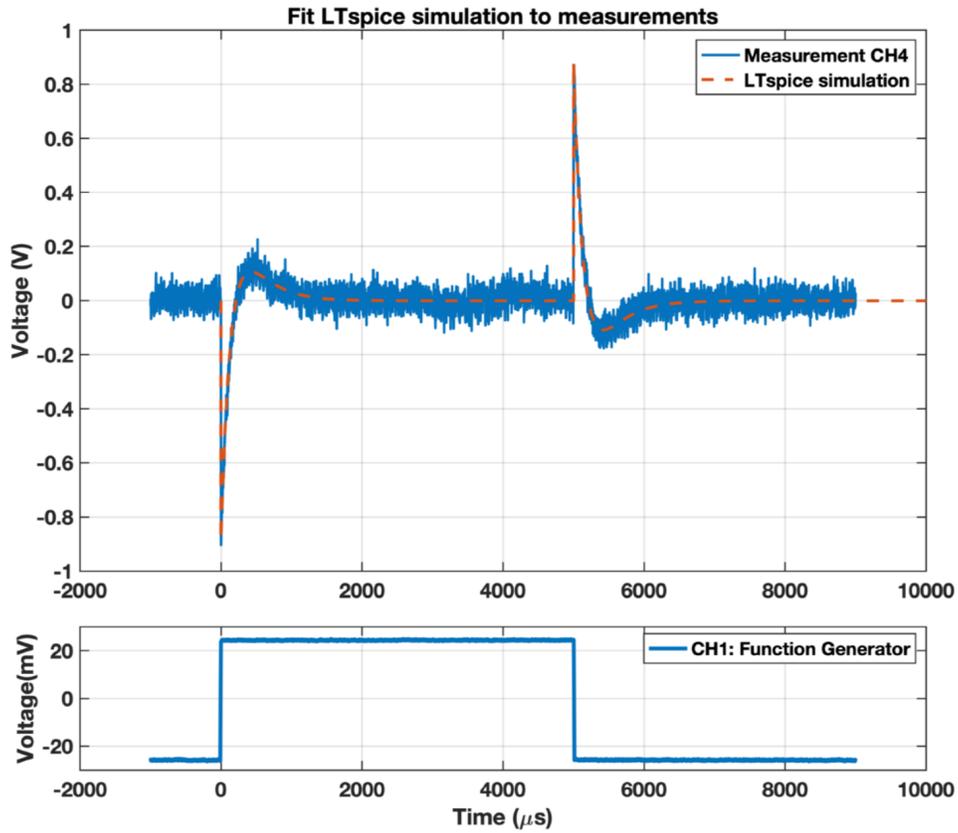

**Fig. A1:** The waveforms of the Function Generator output (bottom panel) and the recorded signal output from the model SC after injecting the test charge onto one of the antennas. See text for detail.

# Acknowledgement

The contributions to this study from authors M.S. and Z.S. were supported by NASA's Cassini Data Analysis Program (CDAP), Grant NNX17AF99G. The contribution from author A.G. was supported by the Europe-Colorado Mobility Program of the College of Engineering and Applied Sciences of the University of Colorado at Boulder. The contribution from author D.M. was supported by NASA contract NNN06AA01C. The operation of the dust accelerator was supported by NASA's Solar System Exploration Research Virtual Institute (SSERVI) Cooperative Agreement Notice, Grant 80NSSC19M0217. The authors thank John Fontanese for operating the dust accelerator during the experimental campaigns. The authors appreciate Craig Joy from the physics department supporting the machining of the spherical spacecraft model. The laboratory data (Shen et al., 2021, supplement dataset) are publicly available in Zenodo repository (http://doi.org/10.5281/zenodo.4888925).

# Author contributions

Conceptualization – Z. Sternovsky

Formal analysis – M. Shen

Funding acquisition - Z. Sternovsky, D. Malaspina

Investigation – M. Shen, A. Garzelli

Methodology – M. Shen, Z. Sternovsky, A. Garzelli

Validation – M. Shen, Z. Sternovsky, D. Malaspina

Writing – original draft –  Z. Sternovsky, M. Shen

Writing – review & editing – Z. Sternovsky, M. Shen, D. Malaspina